\documentclass[conference]{IEEEtran}
\IEEEoverridecommandlockouts
\author{\IEEEauthorblockN{Anonymous Authors}}
\usepackage{cite}
\usepackage{amsmath,amssymb,amsfonts}
\usepackage{graphicx}
\usepackage{textcomp}
\usepackage{xcolor}
\def\BibTeX{{\rm B\kern-.05em{\sc i\kern-.025em b}\kern-.08em
		T\kern-.1667em\lower.7ex\hbox{E}\kern-.125emX}}
\usepackage{amsmath,amsfonts}
\usepackage{array}
\usepackage[caption=false,font=normalsize,labelfont=sf,textfont=sf]{subfig}
\usepackage{textcomp}
\usepackage{stfloats}
\usepackage{url}
\usepackage{verbatim}
\usepackage{graphicx}
\usepackage[export]{adjustbox}
\usepackage{ntheorem}
\usepackage{longtable}
\usepackage{booktabs}
\usepackage{makecell}
\usepackage{amsmath}
\usepackage{graphicx}
\usepackage{mwe}
\usepackage{graphbox}
\usepackage{amsfonts} 
\usepackage{enumitem}
\usepackage{amssymb}
\usepackage{ragged2e}
\usepackage{multicol} 

\usepackage[colorlinks, linkcolor=blue, anchorcolor=blue, citecolor=blue]{hyperref}
\usepackage[numbers,sort&compress]{natbib}
\usepackage{comment}
\usepackage{color}
\usepackage{bbding} 
\usepackage{diagbox} 
\usepackage{stfloats}
\usepackage{float}
\usepackage{multirow}
\usepackage{subcaption}
\theorembodyfont{\upshape}

\newtheorem{theorem}{Theorem}[section] 

\newtheorem{definition}[theorem]{Definition}

\theoremstyle{remark}
\makeatletter
\renewtheoremstyle{plain}
{\item{\theorem@headerfont ##1\ ##2\theorem@separator}~}
{\item{\theorem@headerfont ##1\ ##2\ (##3)\theorem@separator}~}
\makeatother
\def\BibTeX{{\rm B\kern-.05em{\sc i\kern-.025em b}\kern-.08em
		T\kern-.1667em\lower.7ex\hbox{E}\kern-.125emX}}
\usepackage{balance}

\usepackage{lipsum}

\makeatletter
\newenvironment{breakablealgorithm}
{
		\begin{center}
			\refstepcounter{algorithm}
			\hrule height.8pt depth0pt \kern2pt
			\renewcommand{\caption}[2][\relax]{
				{\raggedright\textbf{\ALG@name~\thealgorithm} ##2\par}%
				\ifx\relax##1\relax 
				\addcontentsline{loa}{algorithm}{\protect\numberline{\thealgorithm}##2}%
				\else 
				\addcontentsline{loa}{algorithm}{\protect\numberline{\thealgorithm}##1}%
				\fi
				\kern2pt\hrule\kern2pt
			}
		}{
		\kern2pt\hrule\relax
	\end{center}
}
\makeatother
\usepackage{setspace}
\graphicspath{{QAHAN/EPS/}}
\usepackage{enumitem}
\setenumerate[1]{itemsep=0pt,partopsep=0pt,parsep=\parskip,topsep=0pt}
\setitemize[1]{itemsep=0pt,partopsep=0pt,parsep=\parskip,topsep=0pt}
\setdescription{itemsep=0pt,partopsep=0pt,parsep=\parskip,topsep=0pt}

\usepackage{algorithm} 
\usepackage{algpseudocode} 

\usepackage{algorithmicx}  
\usepackage{algpseudocode}  

\begin{document}

\title{QAHAN: A Quantum Annealing Hard Attention Network 

}

\DeclareRobustCommand*{\IEEEauthorrefmark}[1]{%
  \raisebox{0pt}[0pt][0pt]{\textsuperscript{\footnotesize #1}}%
}

\author{
	\IEEEauthorblockN{
		Ren-Xin Zhao\IEEEauthorrefmark{1,2}, 
	}
	
	\IEEEauthorblockA{\IEEEauthorrefmark{1}School of Computer Science and Engineering, Central South Univerisity, Changsha, China}
	\IEEEauthorblockA{\IEEEauthorrefmark{2}School of Statistics and Computer Science, Trinity College Dublin, Dublin, Ireland}
	
	\IEEEauthorblockA{
		Email: renxin\_zhao@alu.hdu.edu.cn, 
	}
}

\maketitle

\begin{abstract}
Hard Attention Mechanisms (HAMs) effectively filter essential information discretely and significantly boost the performance of machine learning models on large datasets. Nevertheless, they confront the challenge of non-differentiability, which raises the risk of convergence to a local optimum. Quantum Annealing (QA) is expected to solve the above dilemma. We propose a Quantum Annealing Hard Attention Mechanism (QAHAM) for faster convergence to the global optimum without the need to compute gradients by exploiting the quantum tunneling effect. Based on the above theory, we construct a Quantum Annealing Hard Attention Network (QAHAN) on D-Wave and Pytorch platforms for MNIST and CIFAR-10 multi-classification. Experimental results indicate that the QAHAN converges faster, exhibits smoother accuracy and loss curves, and demonstrates superior noise robustness compared to two traditional HAMs. Predictably, our scheme accelerates the convergence between the fields of quantum algorithms and machine learning, while advancing the field of quantum machine vision.
\end{abstract}


\section{Introduction}\label{introduction}

\IEEEPARstart{H}{AM}s which are modules that reduce the computational burden and improve the modeling efficiency by sparsely filtering the important information when dealing with the large-scale data \cite{0.0,0.1,0.2,0.3}, have achieved tremendous success in areas such as computer vision \cite{0.4} and natural language processing \cite{0.5}. One of the striking cases is that a HAM achieves two impressive accuracies of 95.83\% and 99.07\% for safe driving recognition and driver distraction detection, respectively, in addition to a 38.71\% reduction in runtime \cite{0.6}.
Albeit efficient, a large portion of HAMs are difficult to optimize continuously by traditional gradient descent methods due to discrete sparse selection \cite{0.7}. This discretization causes the optimization to lack smoothness, making it difficult to efficiently convey the gradient information, which triggers vanishing or unstable gradients. It also raises the risk of falling into the local optima, which ultimately impacts the model performance and makes the optimization process slow and unreliable. These issues are especially prominent in complex tasks. Moreover, in high-dimensional search spaces, HAMs dramatically extend the training time and storage Requirements, further exacerbating the optimization difficulty and limiting large-scale applications \cite{0.8}. Therefore, addressing the above challenges has become increasingly urgent. Fortunately, the QA shows a promising way for the above dilemmas.

The QA is a quantum optimization algorithm that emerged with the development of the quantum technology \cite{0.8.1,0.8.2,0.8.3,0.8.4,0.8.5}. It exploits the quantum tunneling effect \cite{0.9} to efficiently find global optimal solutions in complex search spaces, which is popular in areas like transportation \cite{0.10} and investment \cite{0.11}. In 1998, the QA was first proposed as a theoretical exploration \cite{0.12}. In 2011, D-Wave Systems launched the first commercial quantum annealer, symbolizing the transition of the QA from theory to application \cite{0.13}. In 2019, the hardware resources of quantum annealers surpassed 5000 qubits \cite{0.14}, which makes the potential benefits of the QA in machine learning increasingly apparent. Here the advantages of the QA are reflected in many ways. For example, the QA was shown to be physically interpretable in the Higgs optimization problem and outperformed traditional machine learning methods on small training datasets in 2017 \cite{0.15}. In 2020, the QA combined with deep neural networks broke the small dataset curse and achieved competitive results in unsupervised and semi-supervised tasks on large datasets \cite{0.16}. In the same year, the application of the QA significantly reduced the classification error from 50\% to 0.6\% for linear classifiers on synthetic datasets and from 4.4\% to 1.6\% on other datasets \cite{0.17}. In 2023, the QA maintained comparable accuracy with further acceleration in the single-image super-resolution problem \cite{0.18}. The above success cases of the QA have profoundly inspired the innovation of this paper:

\begin{itemize}
	\item  A QAHAM is proposed that combines the merit of efficient global search of the QA and the information extraction of the HAM.
	\item  A QAHAN is constructed to filter the important parts of image data to improve the model performance.
	\item Comparative experiments demonstrate that QAHAN has faster convergence speed, higher learning accuracy, lower loss value, and stronger noise robustness than the other two HAMs.
\end{itemize}

The rest of the paper is structured as follows: in Section \ref{sec2}, the HAM and the QA are outlined to provide a theoretical foundation for the creation of the QAHAM. In Section \ref{sec3}, a QAHAM is described in mathematical detail and the corresponding QAHAN is constructed. In Section \ref{sec4}, three experiments are conducted, and a series of meaningful findings are obtained. Finally, conclusions are drawn based on the findings in the previous sections.

\section{Preliminaries}\label{sec2}
In this section, the mechanics of HAM and QA are briefly reviewed.

\subsection{Hard Attention Mechanism}\label{Aa}

The HAM \cite{0.1,0.101} assigns binary weights (0 or 1) to control which features are deemed important and which can be ignored. Its concise principle is as follows.
Let the input be 
\begin{equation}\label{In}
	\setlength{\abovedisplayskip}{3pt}\setlength{\belowdisplayskip}{3pt}
	\mathbf{In}=[{{\mathbf{v}}_{a}}]_{a=1}^{n},
\end{equation}
where ${{\mathbf{v}}_{a}}$ can be a scalar or a vector representing the $i$-th element in the input. The HAM generates a binary mask $\mathbf{x}=[x_a]_{a=1}^{n}$ to selectively focus on input features, where
\begin{equation}\label{selct}
	\setlength{\abovedisplayskip}{3pt}\setlength{\belowdisplayskip}{3pt}
	x_{a}=\left\{ \begin{array}{*{35}{l}}
		1, & \text{if feature} ~{{\mathbf{v}}_{a}}~ \text{is selected}  \\
		0, & \text{otherwise}  \\
	\end{array} \right..
\end{equation}
Then the final output feature vector is
\begin{equation}\label{HAM}
	\setlength{\abovedisplayskip}{3pt}\setlength{\belowdisplayskip}{3pt}
	\mathbf{Out}(\mathbf{x},\mathbf{In})=\mathbf{x}\odot \mathbf{In}=\sum\limits_{\underline{}a=1}^{n}{{x_a}{{\mathbf{v}}_{a}}},
\end{equation}
where $\odot$ denotes element-wise multiplication, ensuring only features corresponding to $x_a=1$ are retained in the output, while those with $x_a=0$ are ignored. Furthermore, note that the treatment of  \eqref{HAM} is not unique. That is, it is not necessary to perform a summation at the end, but stops at filtering the important elements \cite{1.1}.

\subsection{Quantum Annealing}\label{Quantum Annealing}

QA \cite{1.2} typically maps a real-world problem to a Hamiltonian 
\begin{equation}\label{HP}
	\setlength{\abovedisplayskip}{3pt}\setlength{\belowdisplayskip}{3pt}
	\begin{aligned}[b]
		{{H}_{P}}&={{\mathbf{x}}^{\text{T}}}\mathbf{Qx} \\ 
		& =\left[ \begin{matrix}
			{{x}_{1}} & \cdots  & {{x}_{n}}  \\
		\end{matrix} \right]\left[ \begin{matrix}
			{{q}_{11}} & \ldots  & {{q}_{1n}}  \\
			\vdots  & \ddots  & \vdots   \\
			{{q}_{n1}} & \cdots  & {{q}_{nn}}  \\
		\end{matrix} \right]\left[ \begin{matrix}
			{{x}_{1}}  \\
			\vdots   \\
			{{x}_{n}}  \\
		\end{matrix} \right] \\ 
		& =\sum\limits_{i}{{{q}_{ii}}{{x}_{i}}}+\sum\limits_{i<j}{2{{q}_{ij}}{{x}_{i}}{{x}_{j}}}  ,
	\end{aligned}
\end{equation}
where $x_i\in \{0,1\}$ with $i\in \{1,\cdots ,n\}$. $\mathbf{Q}$ is a symmetric matrix, as $q_{ij} = q_{ji}$ when $i\ne j$. ${q}_{ii}$ and ${q}_{ij}$ are coefficients that encode the problem constraints and interactions. The system initially starts in the ground state of a simple Hamiltonian 
\begin{equation}\label{H0}
	\setlength{\abovedisplayskip}{3pt}\setlength{\belowdisplayskip}{3pt}
	{{H}_{0}}=-\sum\limits_{i}{\delta _{i}^{x}},
\end{equation}
where $\delta _{i}^{x}$ is the Pauli-X operator acting on the $i$-th qubit.
The process of annealing is governed by a time-dependent Hamiltonian 
\begin{equation}\label{Ht}
	\setlength{\abovedisplayskip}{3pt}\setlength{\belowdisplayskip}{3pt}
	H(t)=A(t){{H}_{0}}+B(t){{H}_{P}},
\end{equation}
where $A(t)$ and $B(t)$ smoothly vary over time, ensuring a gradual transition from $H_0$ and $H_P$, enabling quantum tunneling to escape local minima. By the end of the annealing process, the system ideally converges to the ground state of $H_P$, which corresponds to the optimal solution.

\section{Quantum Annealing Hard Attention Mechanism and Network}\label{sec3}

In this section, the operation mechanism of QAHAM and the framework of QAHAN are presented.

\subsection{Quantum Annealing Hard Attention Mechanism}
QAHAM is more efficient in finding the global optimum based on quantum tunneling effect. It is defined as follows.

\begin{definition}[Quantum Annealing Hard Attention Mechanism]
\begin{equation}\label{QAHAM}
	\setlength{\abovedisplayskip}{3pt}\setlength{\belowdisplayskip}{3pt}
\begin{aligned}[b]
   \text{QAHAM}&:={{\mathbf{x}}^{\text{T}}}\mathbf{Qx}+{{\lambda }_{1}}{{\left( \sum\limits_{a=1}^{n}{{{x}_{a}}-k} \right)}^{2}} \\ 
 & +{{\lambda }_{2}}\sum\limits_{a=1}^{n-1}{{{x}_{a}}{{x}_{a+1}}}.  
\end{aligned}
\end{equation}
\end{definition}
 \eqref{QAHAM} consists of three parts, ${{\mathbf{x}}^{\text{T}}}\mathbf{Qx}$, ${{\lambda }_{1}}{{(\sum\nolimits_{a=1}^{n}{{{x}_{a}}-k})}^{2}}$ and ${{\lambda }_{2}}\sum\nolimits_{a=1}^{n-1}{{{x}_{a}}{{x}_{a+1}}}$, for which the derivation is as follows.

Suppose the input is a 2-dimensional matrix which is equally divided into $l\times m$ blocks. Each block is subsequently flattened into a row vector, denoted as
\begin{equation}\label{block}
\setlength{\abovedisplayskip}{3pt}\setlength{\belowdisplayskip}{3pt}
    \{{{\text{B}}_{a}}\}_{a=1}^{l\times m}.
\end{equation}
Then the symmetric matrix
	\begin{equation}\label{Q1}
		\setlength{\abovedisplayskip}{3pt}\setlength{\belowdisplayskip}{3pt}
		\begin{aligned}[b]
   \mathbf{Q} &\triangleq \left[ \begin{matrix}
   {{\text{B}}_{1}}  \\
   \vdots   \\
   {{\text{B}}_{l \times m}}  \\
\end{matrix} \right]\left[ \begin{matrix}
   \text{B}_{1}^{\text{T}} & \cdots  & \text{B}_{l \times m}^{\text{T}}  \\
\end{matrix} \right] \\ 
 & =\left[ \begin{matrix}
   {{\text{B}}_{1}}\text{B}_{1}^{\text{T}} & \ldots  & {{\text{B}}_{1}}\text{B}_{l \times m}^{\text{T}}  \\
   \vdots  & \ddots  & \vdots   \\
   {{\text{B}}_{l \times m}}\text{B}_{1}^{\text{T}} & \cdots  & {{\text{B}}_{n}}\text{B}_{l \times m}^{\text{T}}  \\
\end{matrix} \right]  
\end{aligned}.
	\end{equation}
Obviously, the size of  \eqref{Q1} is closely related to $l$ and $m$. If $l$ and $m$ are too large, the computational complexity and the computational resource consumption for  \eqref{Q1} increase dramatically, hence $l$ and $m$ are chosen prudently. Bringing  \eqref{Q1} into  \eqref{HP}, 
    \begin{equation}\label{HAM1}
    \setlength{\abovedisplayskip}{3pt}\setlength{\belowdisplayskip}{3pt}
        \begin{aligned}[b]
   {{\mathbf{x}}^{\text{T}}}\mathbf{Qx}&={{\mathbf{x}}^{\text{T}}}\left[ \begin{matrix}
   {{\text{B}}_{1}}  \\
   \vdots   \\
   {{\text{B}}_{l \times m}}  \\
\end{matrix} \right]\left[ \begin{matrix}
   \text{B}_{1}^{\text{T}} & \cdots  & \text{B}_{l \times m}^{\text{T}}  \\
\end{matrix} \right]\mathbf{x} \\ 
 & ={{\left( \left[ \begin{matrix}
   \text{B}_{1}^{\text{T}} & \cdots  & \text{B}_{l \times m}^{\text{T}}  \\
\end{matrix} \right]\mathbf{x} \right)}^{2}}  
\end{aligned}
    \end{equation}
    is obtained, where $\left[ \begin{matrix}
   \text{Block}_{1}^{\text{T}} & \cdots  & \text{Block}_{n}^{\text{T}}  \\
\end{matrix} \right]\mathbf{x}$ corresponds to  \eqref{HAM}. $\mathbf{x}$ can be defined as hard attention scores. Thus, we have
\begin{definition}[Quantum Annealing Hard Attention Score Matrix]
    \begin{equation}
      \setlength{\abovedisplayskip}{3pt}\setlength{\belowdisplayskip}{3pt}
\text{QAHASM}:=\mathbf{x}\mathbf{x}^{\text{T}}=\left[ \begin{matrix}
   {{x}_{1}}{{x}_{1}} & \cdots  & {{x}_{1}}{{x}_{l\times m}}  \\
   \vdots  & \ddots  & \vdots   \\
   {{x}_{l\times m}}{{x}_{1}} & \cdots  & {{x}_{l\times m}}{{x}_{l\times m}}  \\
\end{matrix} \right].
    \end{equation}
\end{definition}
Theoretically,  \eqref{HAM1} can be directly used as a QAHAM, but its form is too simple and lacks constraints, which may lead to large continuous areas being assigned a weight of 1, weakening the strength of identification of important data. To avoid the above awkward situation, the sparse penalty term 
\begin{equation}\label{sparse}
	\setlength{\abovedisplayskip}{3pt}\setlength{\belowdisplayskip}{3pt}
	{{\lambda }_{1}}{{\left( \sum\limits_{a=1}^{n}{{x_{{a}}}-k} \right)}^{2}}
\end{equation}
and the adhesion penalty term
\begin{equation}\label{control}
	\setlength{\abovedisplayskip}{3pt}\setlength{\belowdisplayskip}{3pt}
	{{\lambda }_{2}}\sum\limits_{a=1}^{n-1}{{x_{a}}{x_{a+1}}}
\end{equation}
are introduced.  (\ref{sparse}) penalizes deviations from the target $k$ to prevent over-selection of items in $\mathbf{x}$, while  (\ref{control}) restricts two neighboring $x_a$ and $x_{a+1}$ to be selected simultaneously to force the model to select among non-neighboring items. In  \eqref{sparse} and  \eqref{control}, $\lambda_{1}$ and $\lambda_{2}$ are the penalty coefficients. Larger values of $\lambda_{1}$ and $\lambda_{2}$ indicate higher sparsity of global weight 1 and lower probability that all neighboring terms are 1, respectively. $k$ is the number of selected terms.  In this way,  \eqref{HAM1} is transformed into the QAHAM with sparsity in  \eqref{QAHAM}, which can further improve the identification of important information. After that,  \eqref{QAHAM} undergoes a slow evolution in a quantum annealer. Eventually, a set of $\mathbf{x}_{opt}$ that minimizes  \eqref{QAHAM} is searched. Once $\mathbf{x}_{opt}$ is obtained, it is brought into  \eqref{QAHAM}. 
\begin{equation}\label{opt}
	\setlength{\abovedisplayskip}{3pt}\setlength{\belowdisplayskip}{3pt}
    \mathbf{x}_{opt}^{\text{T}}\mathbf{Qx}_{opt}
\end{equation}
is computed to produce the hard-attention processed features. These features are sent to the next layer for further training.


\subsection{Quantum Annealing Hard Attention Network}
  
Based on QAHAM, the framework construction of QAHAN 
\begin{equation}
 \setlength{\abovedisplayskip}{3pt}\setlength{\belowdisplayskip}{3pt}
\text{In}\circ \text{Con}{{\text{v}}_{1}}\circ \text{Con}{{\text{v}}_{2}}\circ \text{QAHAM}\circ \text{FC}\circ \text{OUT}
\end{equation}
is shown in Fig. \ref{framework}, where $\circ$ means the connection between layers.
$\text{In}$ is the input layer. $\text{Con}{{\text{v}}_{1}}$ and $\text{Con}{{\text{v}}_{2}}$ are CCLs for feature extraction. The $\text{QAHAM}$ layer  is used to compute the hard attention score $\mathbf{x}_{opt}$ and the final feature in  \eqref{opt}.
$\text{FC}$ is the fully connected layer with trainable parameters. $\text{OUT}$ is the output layer. In Fig. \ref{framework}, the input data is passed through two CCLs for feature depth extraction. Subsequently, the condensed features are passed through the QAHAM layer for discrete optimization. The QAHAM masked features are then fed into the fully connected layer for parameter optimization. Finally, the above steps are iterated repeatedly to update the parameters until the loss function converges. The specific algorithm of QAHAN is as follows.
\begin{figure}[h]
     \setlength{\abovedisplayskip}{3pt}\setlength{\belowdisplayskip}{3pt}
\centering\includegraphics[width=1\columnwidth]{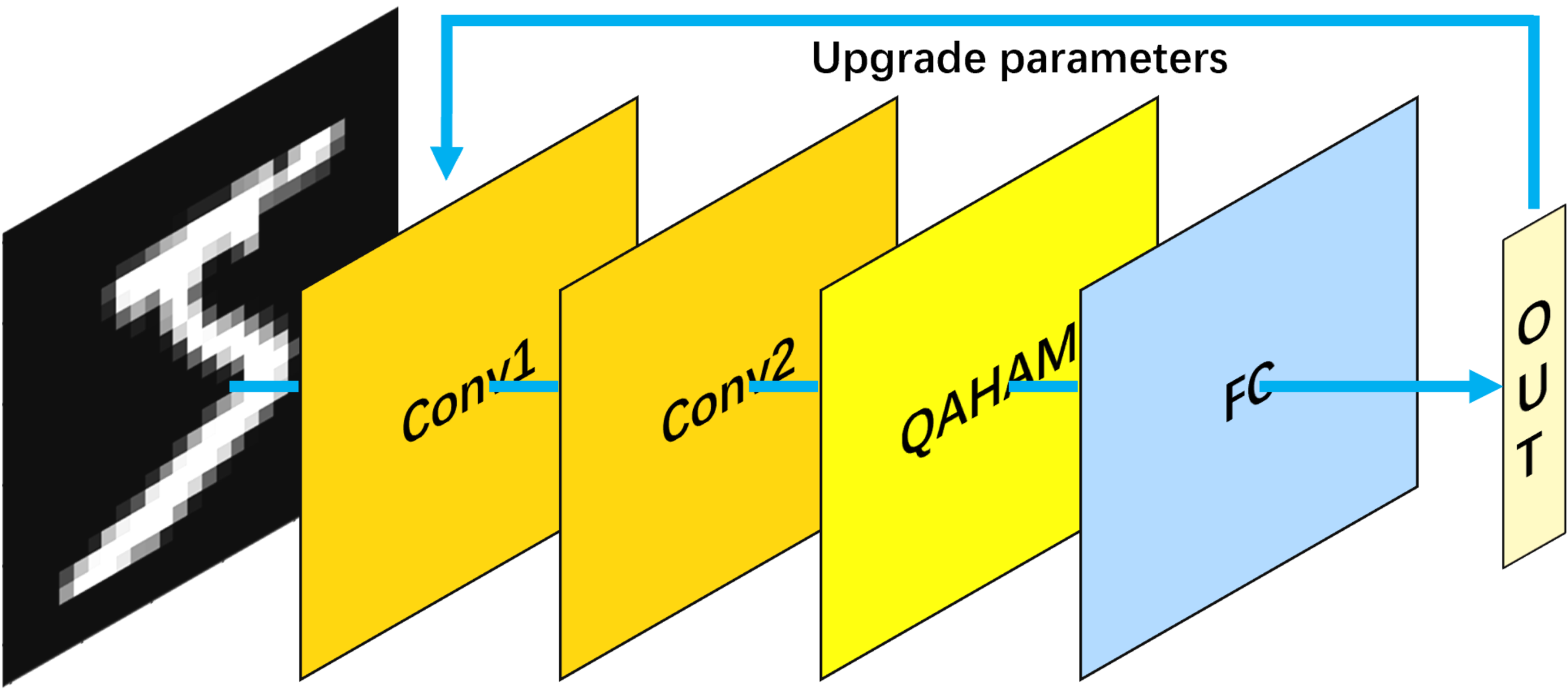}
		\caption{Framework of the QAHAN}\label{framework}
 \end{figure}

\begin{breakablealgorithm}
\caption{QAHAN Algorithm}
\begin{algorithmic}[1]
\Require Dataset, epochs $E$, batch size $B$, selected terms $k$, penalty coefficients $\lambda_1$ and $\lambda_2$
\Ensure Trained QAHAN and training history

\State \textbf{Data Preprocessing:} 
    \begin{enumerate}
        \item Normalize dataset to range $[-1, 1]$.
        \item Subset the dataset to $m$ training samples and $n$ test samples.
        \item Create data loaders with batch size $B$.
    \end{enumerate}

\State \textbf{QAHAM based on D-Wave:}
    \begin{enumerate}
\item Construct  \eqref{Q1} based on  \eqref{block}.
            \item Add  \eqref{sparse} and  \eqref{control}, using selected terms $k$, and penalty coefficients $\lambda_1$ and $\lambda_2$.
            \item Use the D-Wave sampler and perform QA to minimize  \eqref{QAHAM} and determine the optimal binary hard attention score $\mathbf{x}_{opt}$.
            \item Apply $\mathbf{x}_{opt}$ to the input to get  \eqref{opt}.
    \end{enumerate}

\State \textbf{QAHAN Design:}
    \begin{enumerate}
        \item Two convolutional layers with ReLU activation functions.
        \item QAHAM layer applied after the second ReLU and before flattening.
        \item A fully connected layer.
        \item Use dropout for regularization.
    \end{enumerate}

\State \textbf{Training Process:}
    \begin{enumerate}
        \item Initialize optimizer, loss function, and mixed precision scaler.
        \item Initialize training history to record loss and accuracy for each epoch.
        \item For each epoch:
        \begin{enumerate}
            \item Set the model to training mode.
            \item For each mini-batch:
            \begin{enumerate}
                \item Load batch of images and labels.
                \item Forward propagate through the QAHAN model.
                \item Apply QAHAM after the second convolutional layer.
                \item Compute loss and perform backpropagation with gradient scaling.
                \item Update model parameters using the optimizer.
            \end{enumerate}
            \item Compute and store average loss and accuracy for the epoch.
        \end{enumerate}
    \end{enumerate}

\State \textbf{Output:}
    \begin{enumerate}
        \item Trained QAHAN.
        \item Training history containing accuracy and loss for each epoch.
    \end{enumerate}
\end{algorithmic}
\end{breakablealgorithm}

\section{Experiments}\label{sec4}
In this section, the performance of QAHAN is comprehensively evaluated on the D-Wave \cite{1.3} and Pytorch \cite{1.4} platforms to demonstrate the feasibility and effectiveness of our scheme. 
Specifically, the following experiments are conducted. 
\begin{itemize}
    \item QAHAN is performed on MNIST and CIRAR-10 for multi-classification experiments and compared with two other hard attention mechanisms.
    \item Hard attention scores are visualized.
    \item Classification experiments are carried out on the dataset in the presence of weak noise to test the noise robustness of QAHAN
\end{itemize}


\subsection{Dataset and Experimental Setting} 

MNIST \cite{1.5} and CIFAR-10 \cite{1.6} are two widely used benchmark datasets in computer vision research. The MNIST dataset contains 70000 grayscale images of handwritten digits, each of size 28 $\times$ 28 pixels, categorized into 10 classes, primarily used for evaluating image classification algorithms on simple, low-complexity tasks. CIFAR-10 consists of 60000 color images of size 32 $\times$ 32 pixels, spanning 10 categories, suitable for medium-complexity classification tasks. 

The configuration of experiments is described in Tab. \ref{config}. The QAHAM layer is replaced by the HAMs proposed by Mnih \cite{0.0} and Elsayed \cite{0.3} for comparison with QAHAM to highlight the advantages of QAHAN. For the above datasets, 50000 (or 1000) samples are selected as the training (or test) set. The batch size is set as 32 for MNIST or 64 for CIFAR-10. The size of convolutional kernels of Conv$_1$ and Conv$_2$ is $5 \times 5$. The strides and padding of these kernels are 2. The dropout is 0.5 to prevent overfitting of the QAHAN and to improve generalization. FC has 3136 input neurons for MNIST and 4196 for CIFAR-10. This means $l=m=56$ for MNIST and $l=m=64$ for CIFAR-10. The number of output neurons is 10 for both for representing labels.
For the QA, $\lambda_{1}=\lambda_{2}=1.0$, $k=100$. For the optimizer, the Adam optimizer is chosen with a learning rate of 0.001 and a loss function of cross-entropy. The epochs are 70 for MNIST and 100 for CIFAR-10. The QAHASM is initialized with random 0s and 1s at first.
\begin{table}[h]
\setlength{\abovedisplayskip}{3pt}\setlength{\belowdisplayskip}{3pt}
\centering
 \def\tablename{Tab.}
    \caption{Experimental key configurations}
    \label{config}
\resizebox{\columnwidth}{!}{%
\begin{tabular}{@{}lccc@{}}
\toprule
\multirow{2}{*}{indicators} & \multicolumn{3}{c}{models}    \\ \cmidrule(l){2-4} 
                            & QAHAN & Mnih's \cite{0.0}& Elsayed's \cite{0.3} \\ \midrule
training set size      & \multicolumn{3}{c}{50000}          \\ 
test set size      & \multicolumn{3}{c}{1000}          \\ 
batch size      & \multicolumn{3}{c}{MNIST: 32, CIFAR-10: 64}          \\
Conv$_1$     & \multicolumn{3}{c}{kernel: 5 $\times$ 5, stride: 2, padding: 2}          \\ 
Conv$_2$     & \multicolumn{3}{c}{kernel: 5 $\times$ 5, stride: 2, padding: 2}          \\ 
dropout     & \multicolumn{3}{c}{0.5}          \\ 
FC input     & \multicolumn{3}{c}{MNIST: 3136, CIFAR-10: 4096 }          \\ 
FC output     & \multicolumn{3}{c}{10}          \\
QA     & \multicolumn{3}{c}{\(\lambda_1 = 1.0\), \(\lambda_2 = 1.0\), \(k=100\)}          \\
optimizer     & \multicolumn{3}{c}{Adam}          \\
learning rate     & \multicolumn{3}{c}{0.001}          \\
loss function     & \multicolumn{3}{c}{CrossEntropyLoss}          \\
epochs     & \multicolumn{3}{c}{MNIST: 70, CIFAR-10: 100}          \\
initial attention mask     & \multicolumn{3}{c}{random values in \(\{0, 1\}\)}          \\
                      \bottomrule
\end{tabular}%
}
\end{table}

\subsection{Experimental Analysis} 
\begin{figure}[h]
\setlength{\abovedisplayskip}{3pt}\setlength{\belowdisplayskip}{3pt}
\centering\includegraphics[width=1\columnwidth]{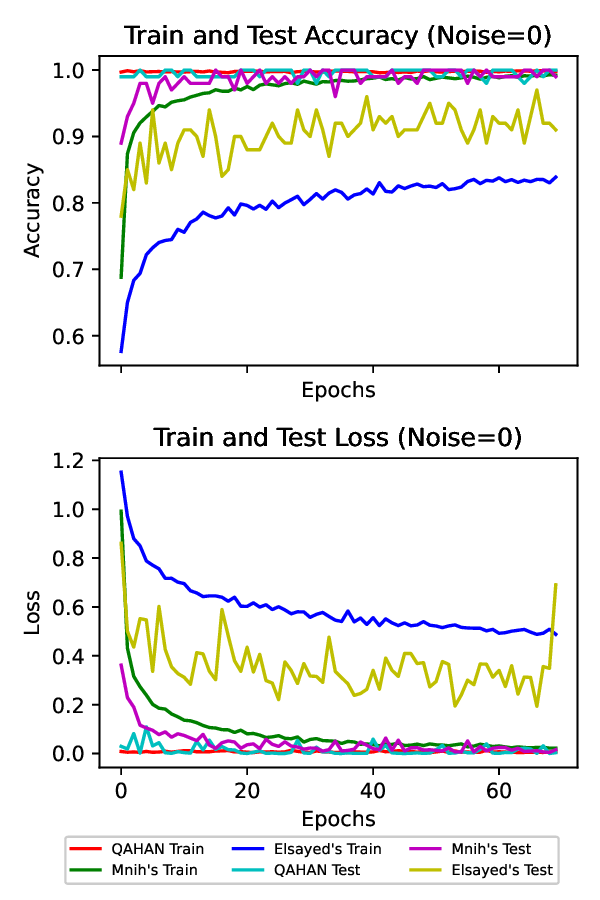}\caption{Noiseless MNIST comparison results}\label{trainMNIST}
\end{figure}
\begin{figure}[h]
\setlength{\abovedisplayskip}{3pt}\setlength{\belowdisplayskip}{3pt}
\centering\includegraphics[width=1\columnwidth]{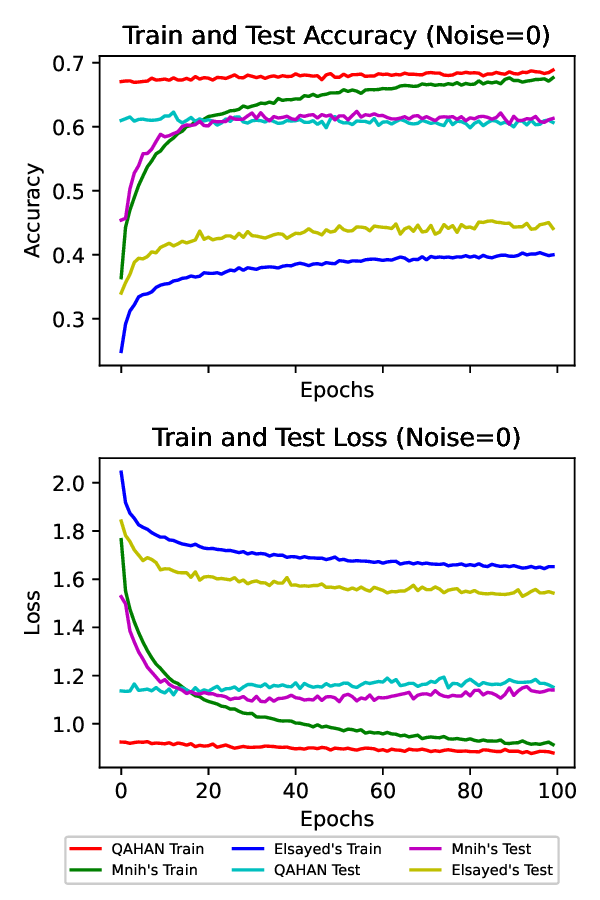}\caption{Noiseless CIFAR-10 comparison results}\label{trainCIFAR}
\end{figure}
\begin{figure}[h]
\setlength{\abovedisplayskip}{3pt}\setlength{\belowdisplayskip}{3pt}
\centering\includegraphics[width=1\columnwidth]{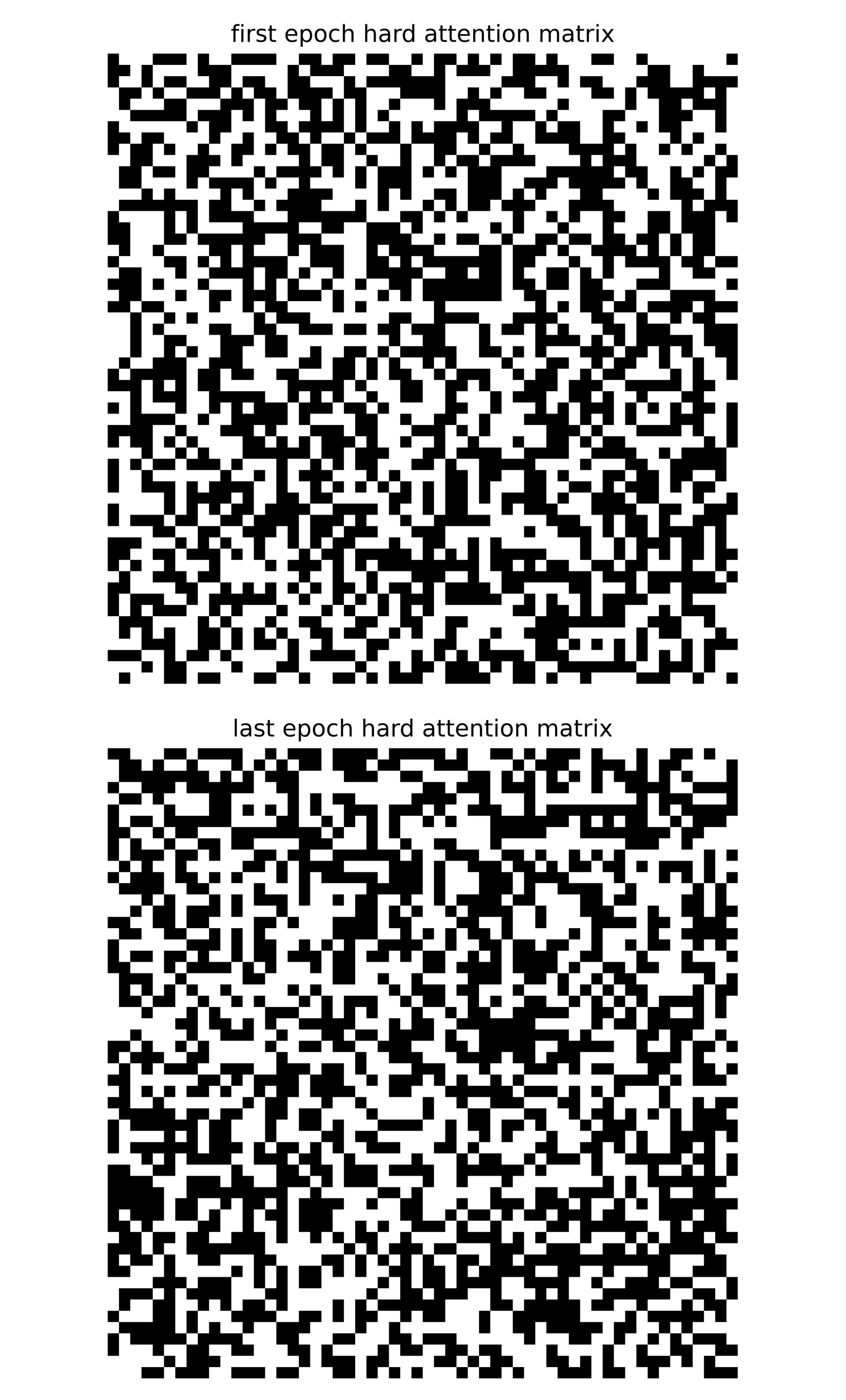}\caption{QAHASM for MNIST}\label{hasMNIST}
\end{figure}
\begin{figure}[h]
\setlength{\abovedisplayskip}{3pt}\setlength{\belowdisplayskip}{3pt}
\centering\includegraphics[width=1\columnwidth]{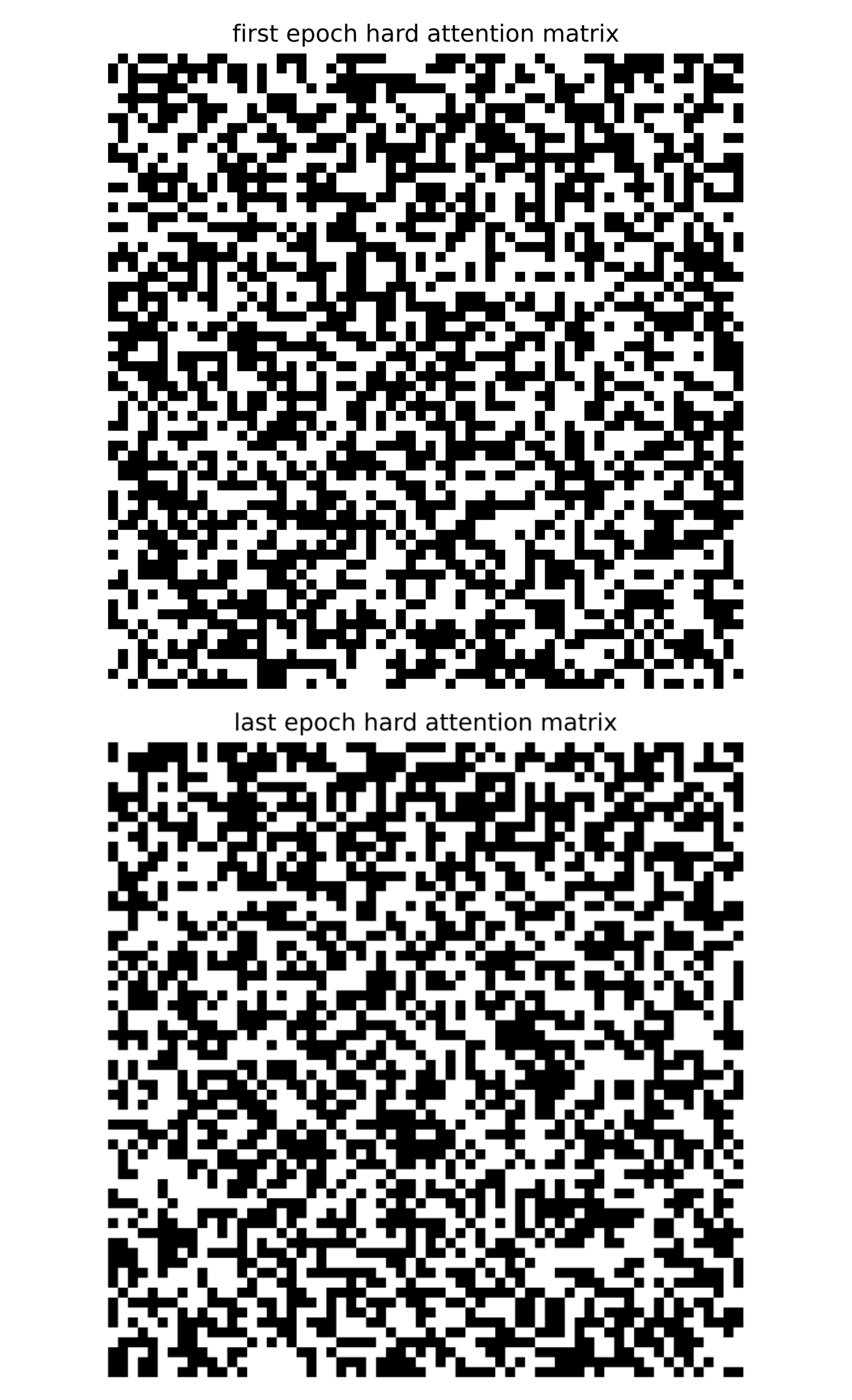}\caption{QAHASM for CIFAR}\label{hasCIFAR}
\end{figure}
\begin{figure}[h]
\setlength{\abovedisplayskip}{3pt}\setlength{\belowdisplayskip}{3pt}
\centering\includegraphics[width=1\columnwidth]{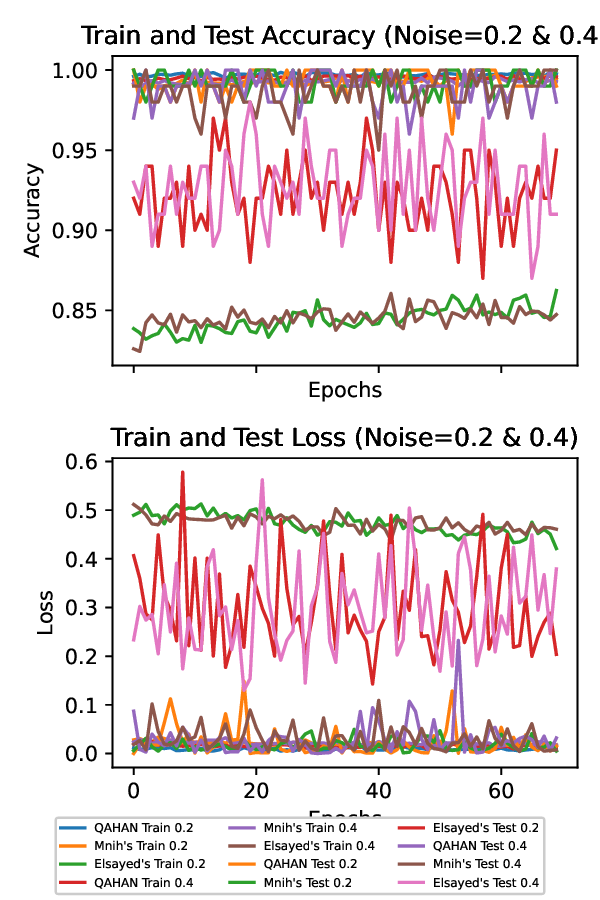}\caption{Noisy MNIST comparison results}\label{noiseMNIST}
\end{figure}
\begin{figure}[h]
 \setlength{\abovedisplayskip}{3pt}\setlength{\belowdisplayskip}{3pt}
\centering\includegraphics[width=1\columnwidth]{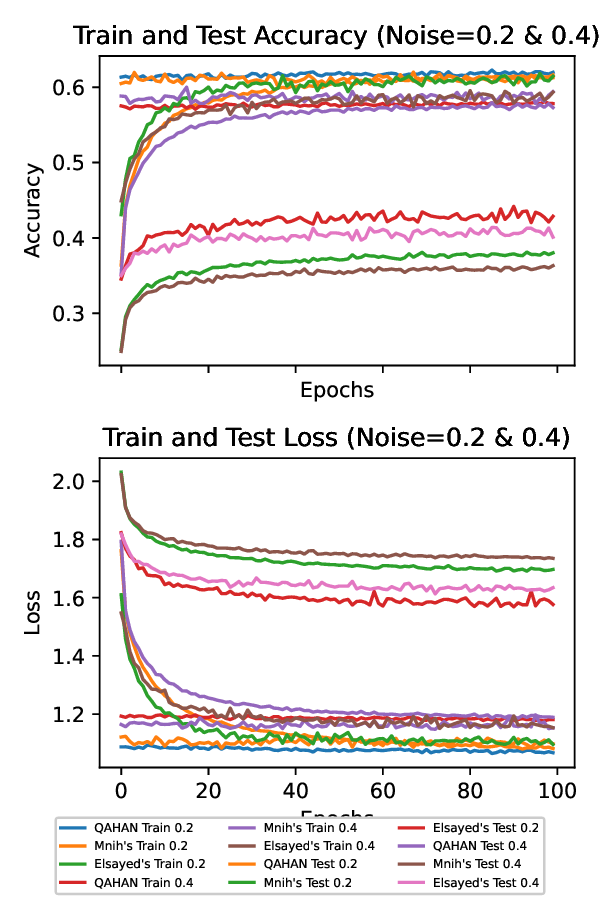}\caption{Noisy CIFAR-10 comparison results}\label{noiseCIFRA}
\end{figure}

\subsubsection{Classification Experiments}
The noiseless comparison results of QAHAN, Mnih's \cite{0.0} and Elsayed's \cite{0.3} methods are shown in Figs. \ref{trainMNIST}  and \ref{trainCIFAR}. In these plots, the horizontal coordinates are all epochs, and the vertical coordinates indicate the accuracies and loss values, respectively. 
By comparison, the conclusions are as follows. 
\begin{itemize}
    \item The QAHAN outperforms Mnih's and Elsayed's models in both training accuracy and testing accuracy. Its training accuracy is stable between 0.996 and 0.999, showing strong data fitting ability and overfitting resistance, while the training accuracies of Mnih's and Elsayed's are lower than 0.95 and 0.85, respectively. In the testing accuracy, QAHAN is always close to 1.0, and especially in the later stages, it is stable between 0.99 and 1.0. In contrast, the Mnih's test accuracy fluctuated between 0.89 and 0.99, and Elsayed's did not exceed 0.94.
    \item Regarding the training loss, the QAHAN loss values are stable and low, usually centred between 0.003 and 0.013. In contrast, the Mnih's and Elsayed's models have higher and more fluctuating training losses, ranging from 0.2 to 0.99 and 0.49 to 1.15, respectively. On the test loss, QAHAN similarly surpasses the other two models with stable loss values between 0.0002 and 0.058, further highlighting its excellent learning performance and loss control.
    \item Compared with Mnih's and Elsayed's models, the QAHAN demonstrates a higher convergence efficiency as its training loss starts to decrease rapidly from the 2nd epoch onwards and is substantially reduced to below 0.0049 after the 5th epoch. Incontrast, Mnih's and Elsayed's models have higher fluctuation of loss in the first 30 epochs, and converge more slowly, especially before the 30th round, the training loss is still high, and fails to stabilise quickly. The test loss of the QAHAN also indicates a significant advantage, especially in the first 20 epochs of the test loss decreases steadily, which is far superior to the other two models, proving that it has a stronger ability to generalise at the early stage of the training period.
\end{itemize}

\subsubsection{Quantum Annealing Hard Attention Score Matrix}
Figs. \ref{hasMNIST} and \ref{hasCIFAR} demonstrate the variation of QAHASM for MNIST and CIFAR-10 in the first and last iterations without noise. The QAHASM in Fig. \ref{hasMNIST} (or \ref{hasCIFAR}) is uniformly divided into  $56\times 56$ (or $64\times 64$) squares, where the black part is used to mask the segmented part of the original image and the white part is activated.
This vividly displays the effect of QAHAN in extracting effective information from the original image.

\subsubsection{Noise Experiments}
The comparison results of QAHAN, Mnih's and Elsayed's models under noise interference are shown in Figs. \ref{noiseMNIST} and \ref{noiseCIFRA}. Based on the results, the following conclusions can be drawn: 
\begin{itemize}
    \item Even under noise interference, the QAHAN consistently outperforms Mnih's and Elsayed's models in accuracies and loss values for both MNIST and CIFAR-10.
    \item The QAHAN reveals a remarkable accuracy robustness under noise perturbation, especially at a noise intensity of 0.2 where the test accuracy stably maintains between 0.98 and 1.0, and at a noise intensity of 0.4 where it stays in the range of 0.96 to 1.0. In addition, the other models fluctuate more in test accuracy under the same noise conditions, while the QAHAN maintains a higher stability. Overall, the QAHAN shows better performance than other models in noisy environments, demonstrating its excellent generalisation ability and noise adaptability.
    \item At a noise intensity of 0.2, the training loss of the QAHAN model is noticeably superior to that of the Mnih and Elsayed models, where the loss values of the QAHAN range from 0.0043 to 0.0185, while those of the Mnih and Elsayed fluctuate from 0.0114 to 0.0464 and 0.1750 to 0.5780, respectively. When the noise intensity is increased to 0.4, QAHAN still displays strong robustness and its loss values remain between 0.0079 and 0.0212, while the loss values of Mnih and Elsayed dramatically increase to 0.0041 to 0.0499 and 0.1769 to 0.5044, respectively. Overall, the QAHAN manifests relatively stable and low loss values in the face of different noise intensities, proving that it suppresses the negative effects of noise notably.
\end{itemize}

\section{Conclusion}\label{sec5}
The HAM has limited performance in high-dimensional systems due to its discrete nature that makes it difficult to find the global optimum efficiently. For this reason, we propose a QAHAM, which combines the advantages of both the QA and the HAM. Based on QAHAM, the corresponding QAHAN is constructed. By comparing the experiments from multiple perspectives, we get the following conclusions: the QAHAN demonstrates a significantly faster convergence rate compared to the Mnih and Elsayed models, with the training loss rapidly decreasing from the 2nd epoch and dropping below 0.0049 after 5th epochs, while the latter models remain unstable even after 30th epochs. Its training and testing accuracies consistently stay within the ranges of 0.996-0.999 and 0.99-1.0, far surpassing the other models, which are below 0.95 and 0.89, respectively. Furthermore, QAHAN maintains a training loss consistently below 0.013, and its testing loss stabilizes between 0.0002 and 0.058, offering a clear advantage over its counterparts. Even under noise interference, QAHAN exhibits remarkable stability in both accuracy and loss values, maintaining testing accuracy between 0.96 and 1.0 and a loss range of 0.0079-0.0212 at a noise intensity of 0.4, showcasing its superior robustness. Overall, QAHAN excels with faster convergence, higher accuracy, lower loss values, and stronger noise resilience.


\begin{thebibliography}{00}
\bibitem{0.0}
V. Mnih et al., “Recurrent models of visual attention,” in Proceedings of the 27th International Conference on Neural Information Processing Systems, pp. 2204–2212, 2014.
\bibitem{0.1}
K. Xu et al., “Show, attend and tell: Neural image caption generation with visual attention,” in Proceedings of the 32nd International Conference on Machine Learning, pp. 2048--2057, 2015.
\bibitem{0.2}
M. Malinowski et al., “Learning visual question answering by bootstrapping hard attention,” in Computer Vision – ECCV 2018, pp. 3-20, 2018.
\bibitem{0.3}
G. F. Elsayed et al., “Saccader: Improving accuracy of hard attention models for vision,” in Proceedings of the 33rd International Conference on Neural Information Processing Systems, pp. 702-714, 2019.
\bibitem{0.4}
D. Wang et al., “Hard attention net for automatic retinal vessel segmentation,” IEEE Journal of Biomedical and Health Informatics, vol. 24, no. 12, pp. 3384-3396, 2020.
\bibitem{0.5}
S. R. Indurthi et al., “Look harder: A neural machine translation model with hard attention,” in Proceedings of the 57th Annual Meeting of the Association for Computational Linguistics, pp. 3037-3043, 2019.
\bibitem{0.6}
I. Jegham et al., “Deep learning-based hard spatial attention for driver in-vehicle action monitoring,” Expert Systems with Applications, vol. 219, pp. 119629, 2023.
\bibitem{0.7}
Z. Niu et al., “A review on the attention mechanism of deep learning,” Neurocomputing, vol. 452, pp. 48-62, 2021.
\bibitem{0.8}
J. Serra, D. Suris et al., “Overcoming catastrophic forgetting with hard attention to the task,” in Proceedings of the 35th International Conference on Machine Learning, pp. 4548-4557, 2018.
\bibitem{0.8.1}
R.-X. Zhao et al., “A review of quantum neural networks: Methods, models, dilemma,” arXiv preprint arXiv:2109.01840, 2021.
\bibitem{0.8.2}
J. Shi et al., “QSAN: A near-term achievable quantum self-attention network,” IEEE Transactions on Neural Networks and Learning Systems, pp. 1-14, 2024.
\bibitem{0.8.3}
R. X. Zhao et al., “QKSAN: A quantum kernel self-attention network,” IEEE Transactions on Pattern Analysis and Machine Intelligence, vol. 46, no. 12, pp. 10184-10195, 2024.
\bibitem{0.8.4}
R.-X. Zhao et al., “GQHAN: A Grover-inspired quantum hard attention network,” arXiv preprint arXiv:2401.14089, 2024.
\bibitem{0.8.5}
R.-X. Zhao et al., “Quantum adjoint convolutional layers for effective data representation,” arXiv preprint arXiv:2404.17378, 2024.
	\bibitem{0.9}
	R. W. Gurney et al., “Wave mechanics and radioactive disintegration,” Nature, vol. 122, no. 3073, pp. 439-439, 1928.
	\bibitem{0.10}
	T. Stollenwerk et al., “Quantum annealing applied to de-conflicting optimal trajectories for air traffic management,” IEEE Transactions on Intelligent Transportation Systems, vol. 21, no. 1, pp. 285-297, 2020.
	\bibitem{0.11}
	E. Grant et al., “Benchmarking quantum annealing controls with portfolio optimization,” Physical Review Applied, vol. 15, no. 1, pp. 014012, 2021.
	\bibitem{0.12}
	T. Kadowaki et al., “Quantum annealing in the transverse Ising model,” Physical Review E, vol. 58, no. 5, pp. 5355-5363, 1998.
	
	\bibitem{0.13}
	M. W. Johnson et al., “Quantum annealing with manufactured spins,” Nature, vol. 473, no. 7346, pp. 194-198, 2011.
	\bibitem{0.14}	
	D. Willsch et al., “Benchmarking advantage and D-Wave 2000Q quantum annealers with exact cover problems,” Quantum Information Processing, vol. 21, no. 4, pp. 141, 2022.
	\bibitem{0.15}
	A. Mott et al., “Solving a Higgs optimization problem with quantum annealing for machine learning,” Nature, vol. 550, no. 7676, pp. 375-379, 2017.
	\bibitem{0.16}
W. Winci et al., “A path towards quantum advantage in training deep generative models with quantum annealers,” Machine Learning: Science and Technology, vol. 1, no. 4, pp. 045028, 2020.
	\bibitem{0.17}
M. Noori et al., “Analog-quantum feature mapping for machine-learning applications,” Physical Review Applied, vol. 14, no. 3, pp. 034034, 2020.
	\bibitem{0.18}
H. Y. Choong et al., “Quantum annealing for single image super-resolution,” in 2023 IEEE/CVF Conference on Computer Vision and Pattern Recognition Workshops (CVPRW), pp. 1150-1159, 2023.
    
	\bibitem{1.1}
    T. Shen et al., “Reinforced self-attention network: A hybrid of hard and soft attention for sequence modeling,” in Proceedings of the 27th International Joint Conference on Artificial Intelligence, pp. 4345–4352, 2018.
    \bibitem{1.2}
E. Pelofske et al., “Short-depth QAOA circuits and quantum annealing on higher-order ising models,” npj Quantum Information, vol. 10, no. 1, pp. 30, 2024.
\bibitem{1.3}
D. Willsch et al., “Benchmarking advantage and D-Wave 2000Q quantum annealers with exact cover problems,” Quantum Information Processing, vol. 21, no. 4, pp. 141, 2022.
\bibitem{1.4}
A. Paszke et al., “Pytorch: An imperative style, high-performance deep learning library,” Advances in Neural Information Processing Systems, vol. 32, 2019.
\bibitem{1.5}
Y. Lecun et al., “Gradient-based learning applied to document recognition,” Proceedings of the IEEE, vol. 86, no. 11, pp. 2278-2324, 1998.
\bibitem{1.6}
B. Recht et al., “Do CIFAR-10 classifiers generalize to CIFAR-10?,” arXiv preprint arXiv:1806.00451, 2018.
\bibitem{1.7}
O. Russakovsky et al., “ImageNet large scale visual recognition Challenge,” International Journal of Computer Vision, vol. 115, no. 3, pp. 211-252, 2015.
\end{thebibliography}
\end{document}